\documentstyle[epsfig]{prhep97}

\def\trhozm{T_\rho^{(2 {m})}}

\def\tkszm{T_{K^{\star}}^{(2 {m})}}
\def\tksdm{T_{K^{\star}}^{(3 {m})}}

\def\sgline{\noalign{\vskip 0.10truecm\hrule\vskip 0.10truecm}}
\makeatletter
\let\chapter\hid@chapter
\makeatother
\begin{document}



\authorrunning{J.H.~K\"uhn, E.~Mirkes and J.~Willibald}
\titlerunning{
Theoretical Aspects 
of $\tau\rightarrow K\pi\pi \nu_\tau$ Decays}
 

\def\talknumber{706} 
\title{
Theoretical Aspects of $\tau\rightarrow K\pi\pi \nu_\tau$ Decays}      
\author{J.H.~K\"uhn, \underline{E.~Mirkes} and J.~Willibald}
\institute{Institut f\"ur Theoretische Teilchenphysik, 
         Universit\"at Karlsruhe,\\ D-76128 Karlsruhe, Germany\\[-10mm]}
\maketitle
\vspace{-4cm}
\hfill \vtop{   \hbox{\bf hep-ph/9712263}
                \hbox{\bf TTP97-53}}\footnote{Talk given 
by E. Mirkes at the International Europhysics Conference on High-Energy Physics
(HEP 97), Jerusalem, Israel, 19-26 Aug 1997. }
\vspace{3.5cm}
\begin{abstract}
Predictions based on the 
chirally normalized vector meson dominance model
for decay rates and distributions of $\tau$ decays
into $K\pi\pi\nu_\tau$ final states are discussed.
Disagreements with  experimental 
results can be traced back to the $K_1$ widths.
\end{abstract}
%
Hadronic $\tau$ decays into final states with kaons
can provide detailed information about low energy hadron physics 
in the strange sector.
Predictions for final states with 2 and 3 meson final states 
\cite{kuehn90,fmkaon}
based on the ``chirally normalized vector meson dominance model''
are in good agreement with recent experimental results \cite{tau96}.
Problems in the
axial vector part in the $K\pi\pi$ final states are
discussed in this contribution.  We believe that these can
be traced back to the $K_1$ widths.

The  matrix element ${\cal{M}}$  for the hadronic $\tau$ decay 
into $K\pi\pi$ final states
$
\tau(l,s)\rightarrow\nu(l^{\prime},s^{\prime})
+K(q_{1},m_{1})+ \pi(q_{2},m_\pi)+ \pi(q_{2},m_{\pi})
$
can be expressed in terms of a leptonic  and a
hadronic  current   as
%
%
$
{\cal{M}}={G}/{\sqrt{2}}\,
\sin\theta_{c}
\,M_{\mu}J^{\mu} 
\label{mdef2h}
$
%
%
with
$
M_{\mu}=
\bar{u}(l^{\prime},s^{\prime})\gamma_{\mu}(1-\gamma_{5})u(l,s) \>.
$
The most general ansatz for the matrix element of the
hadronic  current $J^{\mu}(q_1,q_2,q_3)$  
is characterized by four form factors $F_i$, which 
are in general functions of
$s_1=(q_2+q_3)^2, s_2=(q_1+q_3)^2$, $s_3=(q_1+q_2)^2$
and $Q^2$ (chosen as an additional variable)
\begin{eqnarray}
J^{\mu}&=&T^{\mu\nu}
\left[ \,(q_1-q_3)_{\nu}\,F_1\,+\,(q_2-q_3)_{\nu}\,F_2\,\right]
+\,\,i\,
\epsilon^{\mu\alpha\beta\gamma}q_{1\,\alpha}q_{2\,\beta}q_{3\,\gamma}\,F_3
\label{fidef}
\end{eqnarray}
In Eq.~(\ref{fidef})
$T_{\mu\nu}=  g_{\mu \nu} - (Q_\mu Q_\nu)/Q^2$ denotes a transverse
projector. A possible pseudo-scalar form factor $F_4$ is neglected in 
Eq.~(\ref{fidef}).
The form factors $F_{1}$ and $F_{2} (F_{3})$ originate from the 
$J^P=1^+$ axial vector
hadronic current ($J^P=1^-$ vector current) and correspond to a 
hadronic system in a spin one state.

The resulting choice for the form factors $F_i$ 
for the $\pi^0 \pi^0 K^-$, $K^- \pi^- \pi^+$, $\pi^- \overline{K^0} \pi^ 0$ 
decay modes is summarized by \cite{fmkaon}
\begin{eqnarray}
F^{(abc)}_{1,2}(Q^2,s_2,s_3)\hspace{5mm}&=& \label{f1}
                             {2\sqrt 2 A^{(abc)}\sin \theta_c\over 3f_\pi}
                              G_{1,2}^{(abc)}(Q^2,s_2,s_3) 
                              \\
F^{(abc)}_{3}(Q^2,s_1,s_2,s_3) &=& \label{f3}
                              {A^{(abc)}_3\sin \theta_c
                              \over 2\sqrt 2\pi^2f^3_\pi}  
                              G_3^{(abc)}(Q^2,s_1,s_2,s_3)
\end{eqnarray}
where the Breit-Wigner amplitudes $G_{1,2,3}$ are listed  in
table.~\ref{tabformkpipi}.
The normalization factors are
$A^{(\pi^0 \pi^0 K^-,K^- \pi^- \pi^+,\pi^- \overline{K^0} \pi^ 0)}
     =1/4,-1/2,3/(2\sqrt{2})$
and \\
$A^{(\pi^0 \pi^0 K^-,K^- \pi^- \pi^+,\pi^- \overline{K^0} \pi^ 0)}_3
=1,1,\sqrt{2}.$
\begin{table*}[t]
  \setlength{\tabcolsep}{1.5pc}
  \caption{
Parametrization of the form factors
$F_1$ $F_2$ and $F_3$ in
Eqs.~(\protect\ref{f1},\protect\ref{f3})
for $K\pi\pi$ decay modes.
}
\label{tabformkpipi} 
  \begin{tabular*}{\textwidth}{lll}
  \sgline\sgline
$\begin{array}{c}
channel \\(abc)
\end{array}$ &
$G_1^{(abc)}(Q^2,s_2,s_3)$ &
$G_2^{(abc)}(Q^2,s_1,s_3)$                         \\
\sgline
$\pi^0 \pi^0 K^-$ &
$T_{K_1}^{(a)} (Q^2) \tkszm(s_2)$ &
$T_{K_1}^{(a)} (Q^2) \tkszm(s_1) $
\\[2mm]
$K^- \pi^- \pi^+$ &
$T_{K_1}^{(a)} (Q^2) \tkszm(s_2)$ &
$T_{K_1}^{(b)} (Q^2) T_{\rho}^{(1)}(s_1)$ 
\\[2mm]
$\pi^- \overline{K^0} \pi^ 0$ &
$\begin{array}{l}
\,\,\,\,\frac{2}{3} T_{K_1}^{(b)} (Q^2) \trhozm (s_2) \\[1ex]
+ \frac{1}{3} T_{K_1}^{(a)} (Q^2) \tkszm(s_3) 
\end{array}$ &
$\begin{array}{l}
\frac{1}{3} T_{K_1}^{(a)} (Q^2) \times \\[1ex]
\left[ \tkszm(s_1)
 - \tkszm(s_3) \right]
\end{array}$
\\
  \end{tabular*}
  \begin{tabular*}{\textwidth}{lc}
  \sgline
&
$G_3^{(abc)}(Q^2,s_1,s_2,s_3)$
\\
\sgline
$\pi^0 \pi^0 K^-$ &
$\frac{1}{4} \tksdm(Q^2) \left[ \tkszm(s_1) 
- \tkszm(s_2) \right]$
\\[2mm]
$K^- \pi^- \pi^+$ &
$\frac{1}{2} \tksdm(Q^2) \left[ \trhozm(s_1) 
+  \tkszm(s_2) \right]$
\\[2mm]
$\pi^- \overline{K^0} \pi^ 0$ &
$\frac{1}{4} \tksdm(Q^2) \left[2 \trhozm(s_2) 
+ \tkszm(s_1)
+ \tkszm(s_3) \right]$
\\ 
\sgline
  \end{tabular*}
\end{table*}
The form factors $F_1$ and $F_2$ are governed by the
$J^P=1^+$ three particle resonances with strangeness 
\begin{eqnarray}
   T_{K_1}^{(a)}(Q^2) & = &
 \frac{1}{1 + \xi}
   \Big[ \mbox{BW}_{K_1(1400)}(Q^2) + \xi \mbox{BW}_{K_1(1270)} (Q^2)
   \Big] 
\nonumber \\[2mm] 
   T_{K_1}^{(b)}(Q^2) & = & \mbox{BW}_{K_1(1270)}(Q^2) 
\end{eqnarray}
with $\xi = 0.33$  \cite{fmkaon}.
Here, BW denote
normalized Breit-Wigner propagators 
\begin{equation}
\mbox{BW}_{K_1}[s]\equiv 
\frac{-m^2_{K_1}+im_{K_1}\Gamma_{K_1}}{[s-m^2_{K_1}+im_{K_1}
\Gamma_{K_1}]}
\label{bwc}
\end{equation}
with \cite{RPP96} (all numbers in GeV)
\begin{equation}
\begin{array}{ll}
  m_{K_1}(1400) = 1.402 \hspace{4mm}
& \Gamma_{K_1}(1400)= 0.174 \\
  m_{K_1}(1270) = 1.270 \ 
 & \Gamma_{K_1}(1270)= 0.090 \\
\end{array}
\label{k1para}
\end{equation}
The  three meson  vector resonance 
in the form factor $F_3$, denoted by $\tksdm$,
and the two meson $\rho$ and $K^\star$ resonances, denoted by
$\trhozm$ and $\tkszm$ in table~\ref{tabformkpipi}, 
are discussed in detail in \cite{fmkaon}.

\begin{figure}[tb]
  \centering                
  \mbox{\epsfig{file=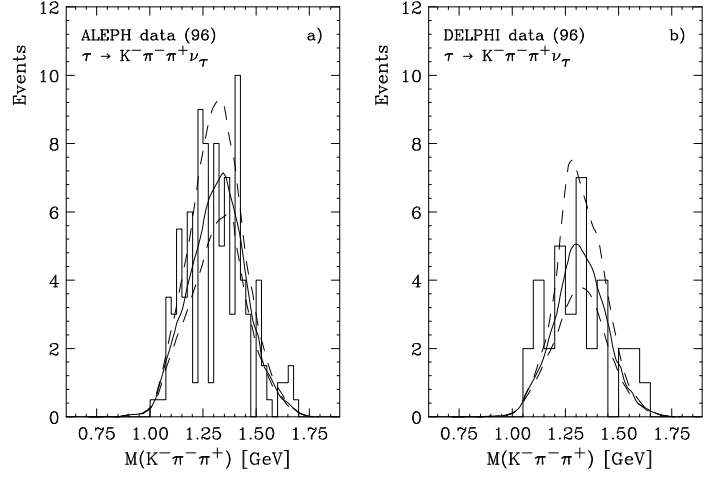,bbllx=0,bblly=240,
                bburx=600,bbury=600,width=0.95\linewidth}} 
 \caption{
Invariant mass $m(K^-\pi^-\pi^+)$   distributions
for the $\tau\rightarrow K^-\pi^- \pi^+\nu_\tau$ decay mode.
The histograms show recent data from (a) ALEPH \protect\cite{aleph}
and (b) DELPHI \protect\cite{delphi}.
The solid line shows the fit result to the $K_1$ widths parameters
in Eq.~(\protect\ref{bwc}) yielding the values in
Eq.~(\protect\ref{fit}).
The dashed lines represent the errors
given in Eq.~(\protect\ref{fit}) for the  $K_1$ widths.
The experimental branching ratios are 0.23 $\pm$ 0.05~\% (ALEPH)
and 0.49 $\pm$ 0.08~\% (DELPHI).
The theoretical predictions for these branching ratios
based on the values in Eq.~(\protect\ref{fit}) are
in good agreement with these numbers.
}
\label{fig1}
\end{figure}

\begin{figure}[tb]
  \centering                
  \mbox{\epsfig{file=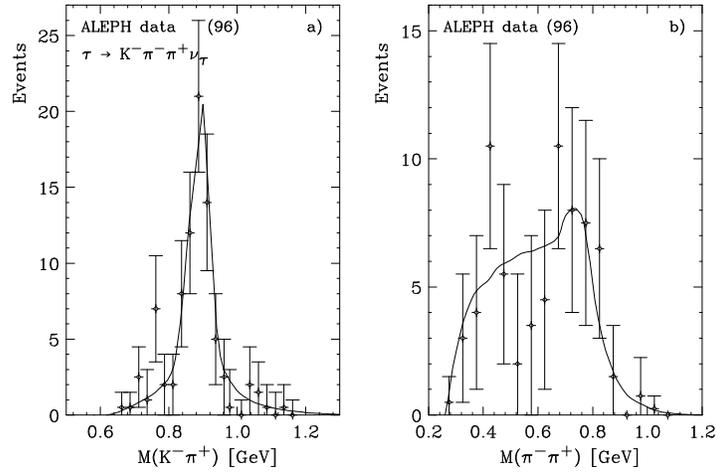,bbllx=0,bblly=240,
                bburx=600,bbury=600,width=0.95\linewidth}} 
 \caption{
$K^-\pi^+$ (a) and $\pi^+\pi^-$ (b)
invariant mass  distributions
for the $\tau\rightarrow K^-\pi^- \pi^+\nu_\tau$ decay mode.
Data are shown from ALEPH \protect\cite{aleph}.
The solid line is the theoretical prediction based
on the $K_1$ parameters  in Eq.~(\protect\ref{fit}).
}
\label{fig2}
\end{figure}

Our predictions 
for the branching ratios of the 
various $K \pi  \pi$ final states based on the above  parameterization
are listed in the second column of table~\ref{tab_kpipi}.
The predictions are considerably larger than
the world averages for the experimental results   presented
at the TAU96 conference (fourth column in table~\ref{tab_kpipi}).
The predictions in the second column in table~\ref{tab_kpipi} are
based on  the particle data group values for the 
widths of the two $K_1$ resonances  (see Eq.~(\ref{k1para})).
We believe that these numbers are considerably too small (see below).
The strong sensitivity of the branching ratios
to the $K_1$ width  is demonstrated by the
numbers  in the third column of table~\ref{tab_kpipi}, where
predictions based on $\Gamma_{K_1}=0.250$ GeV are shown.
The results are now much closer to the measured values.
Our direct fit to recently  measured differential decay distributions for 
the $\tau\rightarrow  K^-\pi^-\pi^+\nu_\tau$ decay mode
by the ALEPH \cite{aleph} and
DELPHI \cite{delphi} collaborations  shown in Fig.~\ref{fig1} yields for the
$K_1$ widths (numbers in GeV):
\begin{equation}
\begin{array}{ll}
\Gamma_{K_1}(1270) = 0.37 \pm 0.1 \hspace{4mm}
& \Gamma_{K_1}(1400)= 0.63 \pm 0.12 \hspace{1cm}\mbox{ALEPH}\\
\Gamma_{K_1}(1270) = 0.19 \pm 0.07  
& \Gamma_{K_1}(1400)= 0.31 \pm 0.08 \hspace{1cm}\mbox{DELPHI}\\
\end{array}
\label{fit}
\end{equation}
with $\chi^2=38/30$ and $\chi^2=15.8/12$, respectively.
The predicted two meson resonance structure based
on these values  is shown in Fig.~\ref{fig2}
and is in good agreement with the experimental data.
\begin{table}[t]
\caption{Predictions for the 
branching ratios ${\cal B}(abc)$ in $\%$ for the
$K\pi\pi$ decay modes.
Results for $K_1$ parameters in Eq.~(\protect\ref{k1para}) 
(second column, vector contribution in parentheses) and for 
$\Gamma_{K_1}(1400)= \Gamma_{K_1}(1270)=0.250 \,\,\mbox{GeV}$ 
(third column) are compared with the experimental world average
(WA) as given at the TAU96 conference (fourth column).}
\label{tab_kpipi}
$$
\begin{array}{c@{\quad}c@{\quad}c@{\quad}c}
\hline \hline \\
\mbox{channel (abc)} &
      \Gamma_{K_1} \,\,\,[\mbox{Eq.}~(\protect\ref{k1para})]&
      \Gamma_{K_1}=0.250 \mbox{GeV} &
       \mbox{WA (TAU96) \protect\cite{tau96}}
 \\ \\
\hline
\\[-2mm]
\pi^0 \pi^0 K^-          & 0.14\,(0.012)  &    0.095   & 0.098\pm 0.021   \\ 
K^-   \pi^-\pi^+         & 0.77\,(0.077)  &    0.45    & 0.228\pm 0.047   \\ 
\pi^-\overline{K^0}\pi^0 & 0.96\,(0.010)  &    0.53    & 0.399\pm 0.048   \\
\hline \hline
\end{array}
$$
\vspace*{-5mm}
\end{table}


%

\end{document}